# Optically Reconfigurable Electrodes for Dielectric Elastomer Actuators

Short Title: Optically Reconfigurable Electrodes

**Authors:** Gino Domel, [1]† Ehsan Hajiesmaili,[1]†‡ David R. Clarke*[1]

**Affiliations:**

[1]John A. Paulson School of Engineering and Applied Sciences, Harvard University; Cambridge, MA 01240, USA.

† These authors contributed equally to this work.

‡ Now at Meta Reality Labs, Redmond, WA 98052.

*Corresponding author.
  Email: clarke@seas.harvard.edu
  Orcid number: 0000-0002-5902-7369

**ABSTRACT:** An optically addressable and configurable electrode architecture for dielectric elastomer actuators and arrays is described. It is based on embedding photoconductive, zinc oxide (ZnO) nanowires in the DEA to create electrodes. Normally, a network of ZnO nanowires is electrically insulating but it becomes conductive in the presence of UV light with a photon energy above the optical bandgap. Taking advantage of this characteristic optical induced switching behavior, we create an optically addressable electrode design, and create new, localized capacitor structures. As the ZnO nanowires are only conductive where, and when, illuminated the effective electrode structure is not fixed, as is the case with CNT and carbon-black electrodes currently used in DEAs. This provides greater, previously unattainable, freedom in the design of dielectric elastomer actuators for soft robotics and devices.

**KEYWORDS:** Optically configurable electrodes. Percolative nanowires. Photoconductive ZnO nanowires. Dielectric elastomer actuators.

# 1. INTRODUCTION

Dielectric elastomer actuators (DEAs) are soft, polymer-based devices that can change shape reversibly when a voltage is applied (*1*). Their large strain and high energy density have led to them being referred to as artificial muscles (*2, 3*). In recent years, much research has been performed to harness this deformation, and a wide range of uses have been found for DEAs in optics, soft robotics and haptics (*4, 5)* including a variety of haptic and sensing applications with soft devices physically attached to human arms and fingers (*6-8*). Several bioinspired soft robotic devices have also been created such as inchworms, caterpillars, and jumping beetles (*9-12*). Furthermore, the use of DEAs has expanded to mimic bioinspired underwater creatures, such as fish, cephalopods, and rays (1*3-15*) and, in the air, as robotic insects (*16, 17*).

These varied applications of DEAs all rely, at their simplest, on a simple compliant capacitor structure consisting of two compliant electrodes separated by a thin dielectric elastomer. When a voltage difference is applied between the two electrodes, they attract each other and exert a compressive force on the elastomer between them. The attractive electrostatic force causes the elastomer to thin as voltage is applied and, because elastomers are nearly incompressible, expand in the other two perpendicular directions to create actuation. It is this actuation, which is reversible with electric field, that is harnessed in many DEA applications. The stress, σ, produced by the electrostatic forces, commonly known as the Maxwell Stress, is proportional to the square of the applied electric field, *E*, where $\sigma = \varepsilon_r \varepsilon_0 E^2$, and $\varepsilon_r$ is the dielectric constant of the elastomer, and $\varepsilon_0$ is the permittivity of a vacuum (*1, 18*).

Recent work has focused on patterning of DEA electrodes and embedded elements in multilayers to give novel shape morphing geometries, with changes in the gaussian curvature to create true shape changes (*19, 20*). These advances have been exploited to create individual arrays of DEAs with shape morphing capabilities (*21*). Nevertheless, although shape morphing DEAs and DEA arrays have significant capabilities, they still behave in a fully deterministic manner that is based on the patterning of the electrode structures, meaning the achievable actuation shape and location of deformation is predetermined by the geometry and stacking of the electrode structure within the DEA.

In this paper, we show that by replacing the CNT electrodes typically used with ZnO nanowires and using local UV illumination, the electrode structure and resulting actuation is not pre-determined by the electrode structure but rather is determined by the location of where the UV excitation is shone on the electrode. This advance builds on the results in a recent publication, in which we have successfully used zinc oxide (ZnO) nanowire channels as interconnects, allowing for optical control of which elements in an array of DEAs are actuated in response to an applied voltage. Due to the absorption and desorption of oxygen on ZnO nanowires when illuminated, ZnO nanowires become electrically conductive in the presence of UV light above its bandgap but returns to its insulating state when the UV is removed, meaning that these channels could be optically controlled (*21, 27-29*). Nevertheless, in that work (*21*), a passive matrix of CNT electrodes was used with the typical cross-talk issues of any passive matrix of capacitors, e.g., addressing two independent spots on the matrix using two pairs of illuminated channels resulted in actuation of four spots on the matrix, two of which being undesired side effects. This hinders the use of such passive matrix architecture for true shape-morphing where addressing multitude of independent locations is required. In this work, we use embedded sheets of ZnO nanowires as electrodes. When illuminated by UV while a voltage is applied, the local electrode structure is modified forming new, ad-hoc, local capacitor structures, and enabling local electric field actuation as well as wiring simplification. Thus, this combination of ZnO nanowires and localized UV illumination allows for reprogrammable, non-contact, and local control of the actuation of the DEA, expanding the potential application of complex DEAs.

In the following sections, the embedded ZnO nanowire networks are shown to have suitable photo-induced electrical conductivity without adversely affecting the stiffness for use as electrodes in DEAs. To demonstrate local control of actuation at the location where the ZnO nanowire electrode is illuminated by UV light, local deflections of single and multilayer sheets are presented. Calibration experiments with optically stimulated ZnO nanowire electrodes located at different depths in elastomer multilayers are described. Then, having shown that ZnO nanowire electrodes can be used in DEA multilayer stacks, local optical control of DEA actuation is demonstrated with four examples of dynamic addressing of UV-controlled DEAs.

## 2. THE OPTICALLY CONFIGURABLE ELECTRODE ARCHITECTURE CONCEPT

The basis of the optically reconfigurable electrode architecture structures is shown schematically in Fig. 1A for a cross-section through a simple parallel plate capacitor structure. Physically, the cross-section consists of a uniform thickness dielectric elastomer with inter-digitated CNT interconnects on both sides but offset and staggered so that they do not overlap in the vertical direction. Both sets of CNT interconnects are covered by a thin layer of ZnO nanowires. When a voltage is applied to the opposing, but offset, CNT interconnects they act as narrow electrodes and a spatially complex electric field distribution in the dielectric is created as illustrated with the electric field lines computed using COMSOL. The net attractive Coulombic force between them is relatively small and spatially varying as seen in Fig. 1B. (For a given elastomer thickness, the net force depends in detail on the electrode offset and their cross-sectional area). When part of the ZnO electrode layers is illuminated with UV light having a photon energy greater than the bandgap of the ZnO (~3.3 eV), the density of the mobile electrons increases, and the illuminated ZnO area becomes electrically conducting. In this way, as shown in Fig. 1C, the illuminated areas of ZnO nanowires become an electrically conducting electrode and, where it overlaps with the CNT interconnects, can be charged to create a local capacitor structure defined by the size of the UV beam and creating a corresponding region that will actuate in response to the local electric field. The field lines are perpendicular to the locally charged electrode structure created by the UV illumination, and the net Coulombic forces are now proportional to the local area illuminated and inversely proportional to the elastomer thickness between the electrode, just as it is in a standard DEA. When the UV illumination is turned off, the ZnO electrode reverts to being highly insulating, the electric field distribution returns to that shown in Fig. 1A, and actuation ceases.

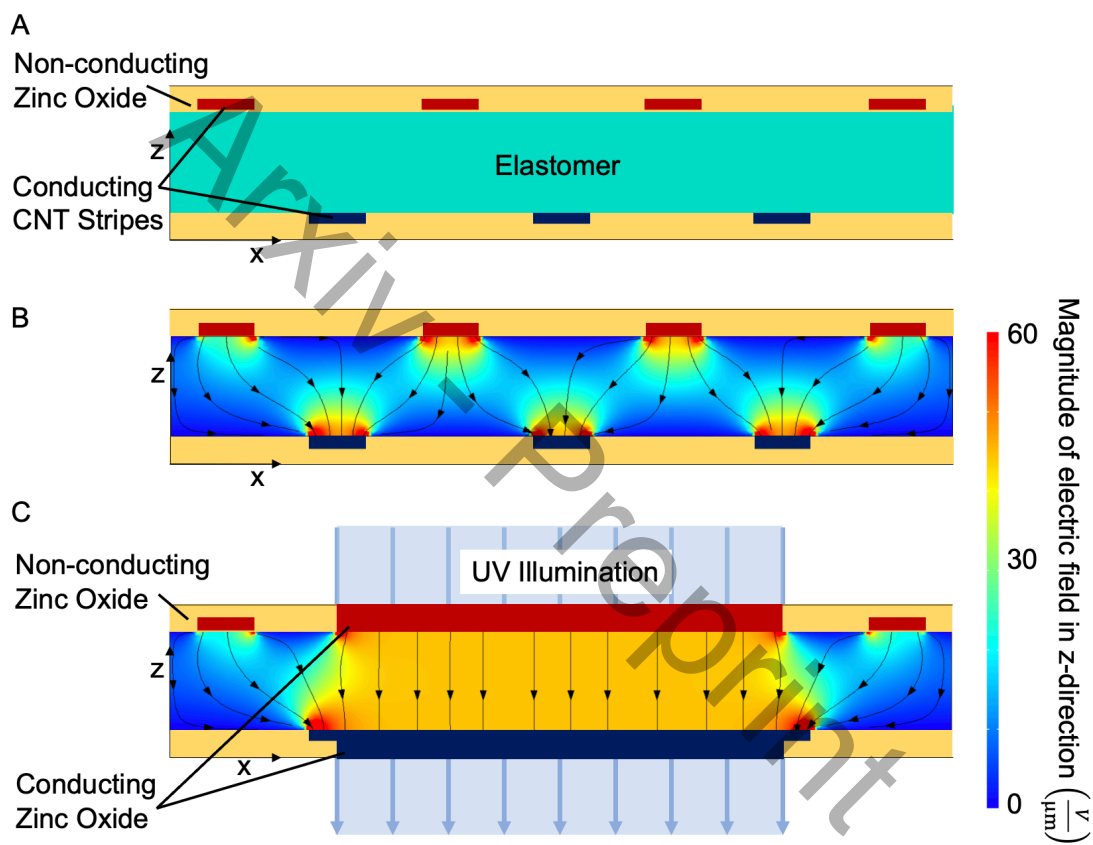

**Fig. 1. Optically reconfigurable electrode design.** (**A**) Cross-section of the electrode structure illustrating a continuous film of ZnO nanowires on either side of an elastomer sheet and interdigitated CNT stripes. Not to scale. (**B**) Electric field distribution in the elastomer when the CNT stripes are connected to a power supply. (**C**) Electric field distribution when the ZnO film is illuminated by UV with an energy exceeding its band-gap. The UV induced conductance locally creates in-situ electrodes and actuation in a dielectric elastomer device. COMSOL simulations of the electric field.

A characteristic feature of the reconfigurable electrode structure is that it is defined by the size and position of the UV illumination and not the CNT interconnects themselves. Consequently, whereas the CNT interconnect structure is fixed in the actuator structure, the actuated regions can be continuously varied over the actuator device by moving the UV source, as shown in a later section and in the videos in Supplementary Materials. This is in marked contrast to the usual role of the CNTs in a DEA to form the electrodes. Here, when used with the ZnO nanowires electrodes, the percolative CNTs become the electrical supply interconnects. In some cases, only two interconnects may be needed, one as the power and the other as ground as will be illustrated with a bimorph fin.

## 3. PHOTOCONDUCTIVITY AND MECHANICAL STIFFNESS OF EMBEDDED ZNO NANOWIRES AT LARGE STRETCHES

The electrodes in dielectric elastomer actuators and sensors must be both electrically conducting and elastically compliant so that they can accommodate the Maxwell stress-induced stretching when a voltage is applied. Both these requirements are usually satisfied in DEA devices by using inherently electrically conducting materials, such as carbon black, carbon nanotubes, or silver nanowires, with concentrations that exceed the percolation threshold up to the maximum stretch of the application. As will be shown, these two requirements can be satisfied with ZnO nanowires that become electrically conducting when illuminated with light having an energy greater than its bandgap.

To assess the change in compliance of different areal densities of ZnO nanowires embedded in elastomer sheets, their tensile stiffness was measured as described in the Supplementary Materials. The effects for three different areal ZnO nanowire densities on the elastic modulus are shown in Fig S1. These results show that the shear modulus (417 kPa) of the CN 9208 elastomer increased successively to 435 kPa, 458 kPa, and 467 kPa as the areal density was increased to 0.25 to 0.5 and then to 1.0 μg/mm$^2$. These increases are smaller than when CNT are used, consistent with the significantly smaller elastic modulus of ZnO (~ 140 GPa) compared to single wall CNTs (~ 1000 GPa).

To determine the density of ZnO nanowire necessary to maintain conductivity even up to large stretches, conductivity measurements were made as a function of stretch. In our previous work, the photoconductivity was measured in unstretched samples (*21*), but the effect of stretching on the ZnO nanowire network was not studied. In this work, the same dog-bone shaped tensile geometry employed for evaluating elastic modulus was used to facilitate both the ease of stretching and making electrical contact. When a voltage was applied between the two CNT regions (as shown in Fig. 2A) in the absence of UV light, no current flowed since the ZnO nanowire bridge was insulating. However, when UV light is shone on the ZnO bridge, the ZnO becomes conductive, allowing current to flow through the sample.

Samples were prepared with three different ZnO nanowire densities (0.08, 0.25, 0.5 μg/mm$^2$) to compare how the density of the ZnO nanowires affects the photoconductance. With 2 kV applied (though automatically modulated if a maximum current is reached), current measurements were recorded with the UV light off for the first 5 seconds, on for 10 seconds, and then off again as indicated by the white and blue shaded regions in the graphs in Fig. 2 (B-D). Current measurements were taken for each sample in its unstretched state and then at incremental stretches of 0.05 from 1.05 up to a maximum of 1.5.

As indicated by the arrows in Fig. 2(B-D), as the stretch was increased, the conductance decreased for all three densities of ZnO. This is as expected since the ZnO nanowires forming the percolative network move relative to one another as the sample is stretched. For the lowest density of 0.08 μg/mm$^2$, even the smallest stretch of 1.05 lead to an order of a magnitude drop in conductance, and at high stretches, the time required to measure any level of conductance with the UV light was as long as 6 seconds.  For the intermediate density (0.25 μg/mm$^2$), only at intermediate and high stretches was there an order of a magnitude drop in conductance and a noticeable rise time in the conductance with the UV light on. This is a significant improvement over the low density (0.08 μg/mm$^2$) ZnO percolative network. For the high density (0.5 μg/mm$^2$), even at the largest stretch of 1.5, there was very little change in UV induced conductance and the response time was less than a second indicating its suitability as a DEA electrode under any intermediate stretch.

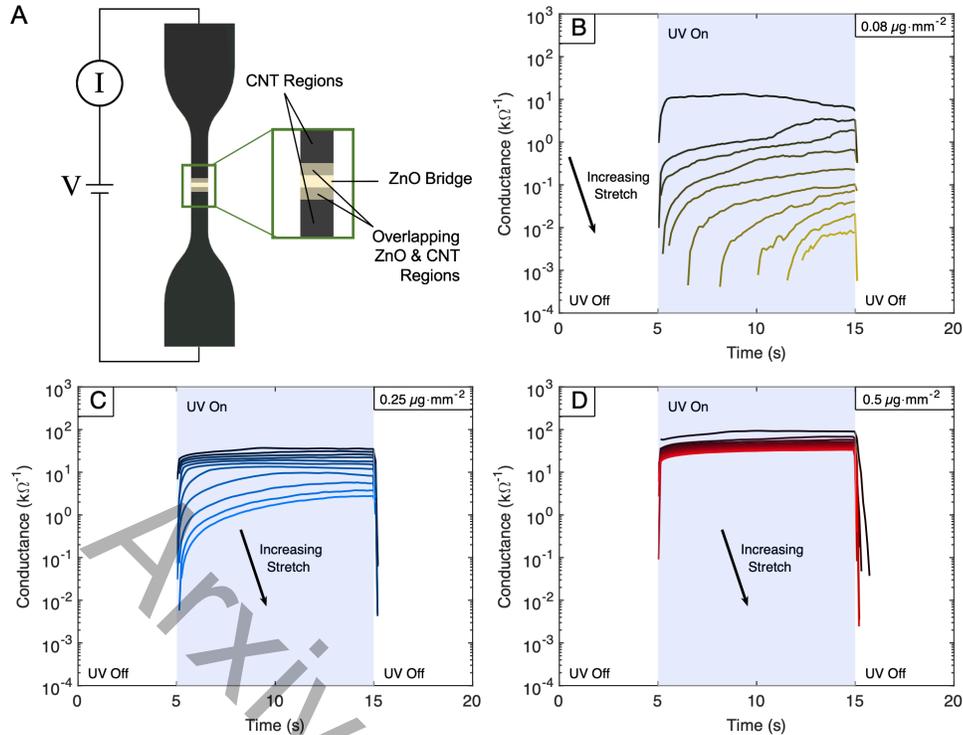

**Fig. 2. Photoconductance measurements as a function of stretch for three ZnO nanowire densities.** (**A**) The test configuration consisted of the elastomer sheet with two CNT electrode regions separated by a 1 mm gap bridged by only ZnO nanowires and elastomer. (**B-D**) A voltage of 2 kV was applied, and a 70 mW/mm$^2$ UV light shone onto the ZnO bridge for ten seconds as indicated by the blue shaded regions. Current measurements were taken first in the unstretched state and then repeated at incremental stretches of 0.05 from 1.05 up to a final stretch of 1.5. Each measurement is represented by a colored curve, from black (corresponding to the unstretched state) and becoming incrementally more brightly colored correspondingly with increasing stretch. The lowest curve in each graph, corresponds to the fully stretched state of 1.5.

## 4. LOCAL OPTICAL ADDRESSING OF SINGLE AND MULTILAYER DEA SHEETS WITH ZNO NANOWIRE ELECTRODES

To demonstrate the ability to optically address and produce local actuation, a simple structure consisting of a taut, square DEA sheet, constrained around its edges, and connected to a voltage supply was adopted (Fig. 3). Where the UV beam was shone on a portion of the DEA, the sheet in that location expands biaxially. However, because of the mechanical constraints around the edges, the expansion can only be accommodated by the DEA sheet deflecting out-of-plane. The deflection depends on the actuation strain, the thickness of the DEA, and its elastic modulus. The size of the square sheet, 45 mm x 45 mm, was chosen to be many times the total thickness, 590 microns of the DEA. Addressing both single layer and multi-layer DEA stackings were evaluated.

The fabrication sequence of a single DEA layer shown in Fig. 3A as described in the Materials and Methods section. The device, 66 mm x 60 mm with a total of twenty-five, 2 mm wide CNT stripes, is shown in Fig. 3B. Based on the photocurrent testing results in figure 2, the density of the ZnO nanowires was chosen to be 0.25 μg/mm$^2$ since stretches of less than 1.1 were expected for this geometry. This particular ZnO density represented a compromise between an ability to maintain photoconductivity while minimizing the bending stiffness of the DEA.

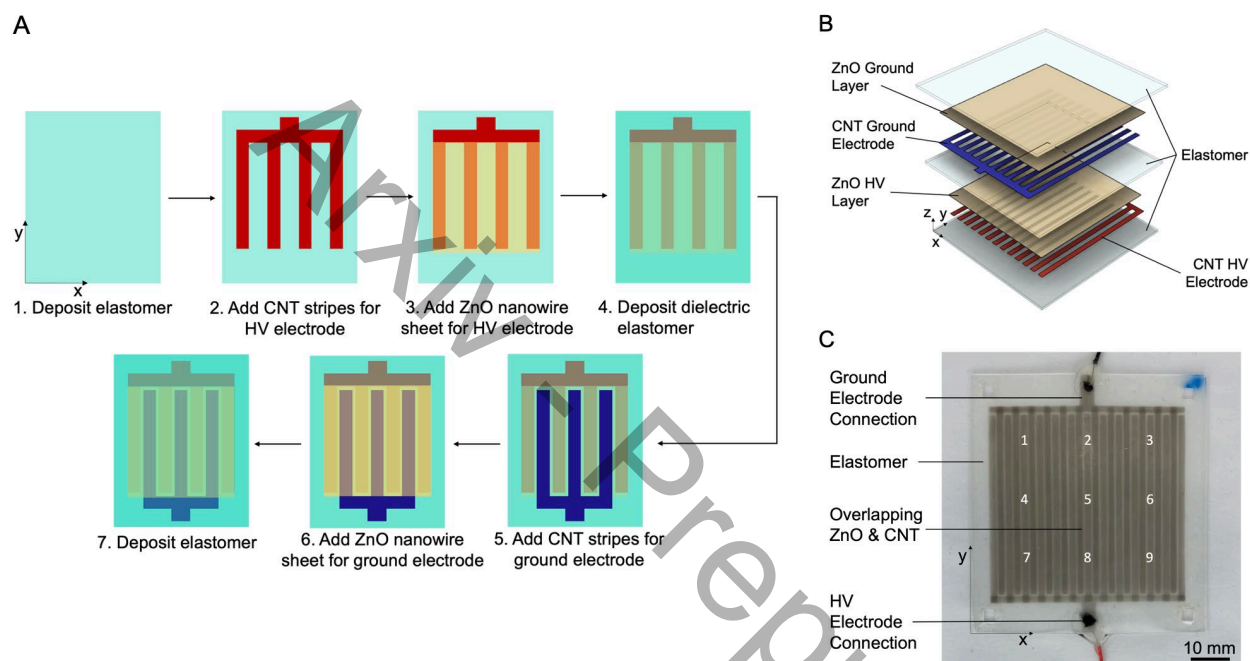

**Fig. 3. DEA fabrication with optically reconfigurable electrode.** (**A**) Fabrication sequence of a simplified, single active layer DEA with a combined ZnO nanowire and CNT electrode architecture. The sequence can be repeated to create a multilayer DEA. Note that HV indicates high voltage. (**B**) Exploded view of the construction of the optically reconfigurable, single active layer DEA with CNT stripes as interconnects and the ZnO nanowire sheets as the electrodes. (**C**) Fabricated DEA. The numbers 1 to 9 indicate the positions illuminated by the UV beam during testing of the local actuation of this device shown in Fig. 5.

In the optical addressing experiments, the UV-produced actuation under an applied voltage, the DEA sheet was clamped into a rigid frame around its four edges with a 45 mm x 45 mm free central region, and the UV source was placed 20 mm away. A mask with nine 5 mm diameter apertures was positioned between the UV source and the DEA sheet and aligned so that UV illumination passes through only one of the apertures in the mask at a time while the other eight holes are covered up. The out-of-plane deflections at nine different locations, where the UV light was shone, and numbered in Fig. 3B, were measured with a laser profilometer. The measurements were repeated for each of the other eight locations (by translating the mask hole through which the UV light was shone). The measured out-of-plane displacements are shown in Fig. S2 and demonstrate that the location of the actuation can be controlled simply by shifting the UV beam.

The experiments were repeated with multilayer DEA sheets since, for a fixed electric field per layer, the actuation force generated increases with the number of layers. However, as both the elastomer and both ZnO nanowires and CNTs absorb in the UV portion of the spectrum, as shown in the optical absorption spectra shown in figure S3, it is necessary to establish the number of layers through which the UV beam can penetrate to create photoconductivity in a ZnO nanowire electrode at different depths below the surface. To assess this, we adopted an empirical approach measuring the actuation deflection of a single DEA layer after the UV beam had first passed through different numbers of DEA layers.

The experimental arrangement used is shown in Fig. 4 and consists of an elastomer screen, covered by a CNT layer and four circular regions having differing concentrations of ZnO nanowires. This multilayer screen was positioned in front of a single layer DEA sheet, the "witness sheet", with off-set interdigitated CNT electrodes on an elastomer with four circular regions having differing ZnO nanowires densities that corresponded to the four densities of the screen. To ensure no displacement occurs other than where the DEA sheet was illuminated, the DEA sheet was clamped at the edges of each of the circular ZnO nanowire regions. A high voltage (2 kV) was applied to the CNT electrodes in the single DEA layer, and the out-of-plane deflections were measured and repeated for different number of layers (each comprised of CNT, ZnO, and elastomer) in the screen placed between the UV source and the DEA sheet.

The results are shown in Fig. 4 (C) for four ZnO nanowire densities and for the number of electroded elastomer layers in the screen. The first data point for each curve in Fig. 4C, corresponds to the single DEA without any intervening screen. The two lower densities (0.08 and 0.25 μg/mm$^2$) showed a larger maximum z-displacement actuation than the two higher densities (0.5 and 1.0 μg/mm$^2$) effectively regardless of the number of layers in the screen, which can be attributed to them adding less stiffness into the soft DEA than the higher two densities. Additionally, the two higher densities showed a noticeable decrease in the maximum z-displacement as the number of layers in the screen approached ten, which is not seen in the curves for the two lower densities. This indicates that the two higher densities of ZnO nanowires are less suitable for multilayer DEAs because of the correspondingly higher absorption of UV by the ZnO nanowire layers. The two lower densities performed similarly well, with 0.25 μg/mm$^2$ exhibiting the highest displacement, further indicating its suitability for multilayers up to at least ten layers based on both its ability to maintain photoconductivity at high stretches (as compared to 0.08

µg/mm$^2$) and allow UV light to reach all layers (as compared to 0.5 and 1.0 µg/mm$^2$), without adding significant unwanted stiffness to the DEA.

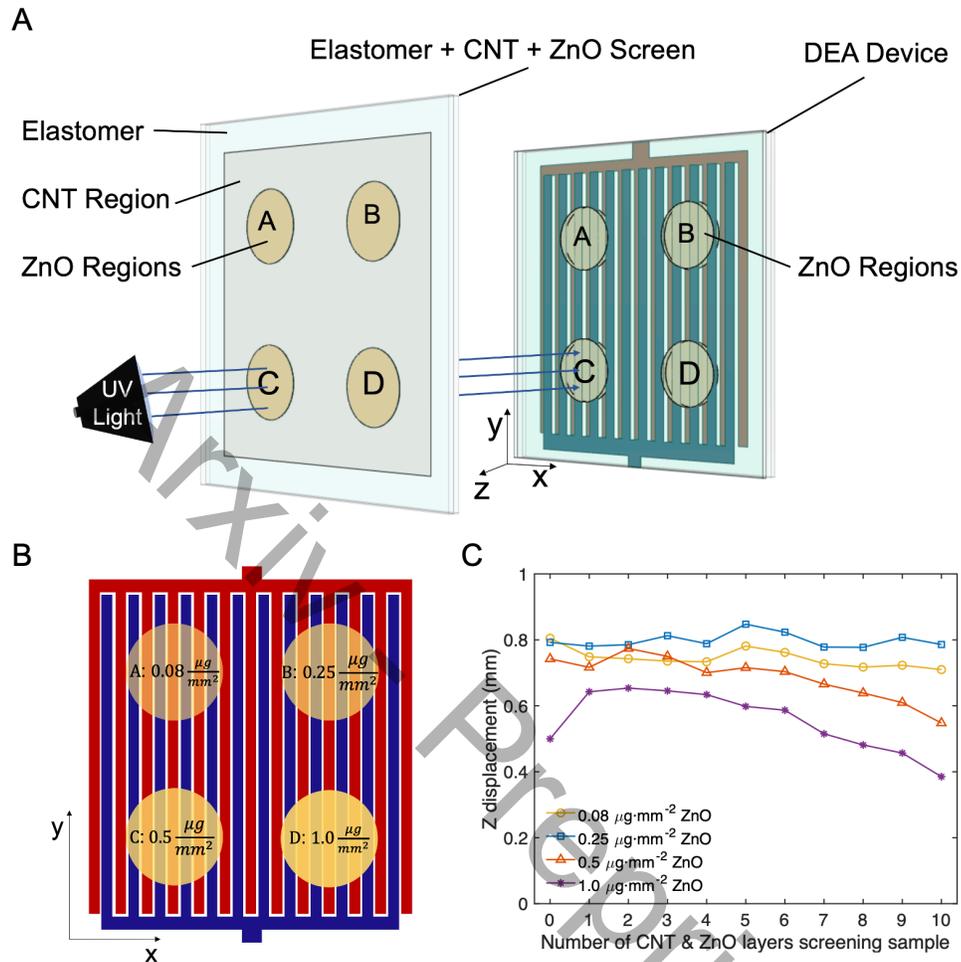

**Fig. 4. Measurement of ZnO nanowire density on multilayer actuation.** (**A**) A single unit DEA device, similar to Fig. 3A, but with four distinct, circular ZnO nanowire regions each with different densities of ZnO nanowires. UV light was shone onto each circular region separately and sequentially, and the maximum z-displacement measured for each ZnO nanowire density. The measurements were then repeated with a screen between the DEA device and the UV light. Each layer of the screen is encapsulated by elastomer and is made up of CNT and a ZnO nanowire region with the same density as the density of the circular ZnO nanowire region lined up directly behind it. After the displacement at each ZnO nanowire density circular region is measured with UV light passing through the screen, another layer of ZnO nanowire, CNT, and elastomer is added to the screen. (**B**) Schematic of electrode architecture with the CNT stripes (in the red and blue) and the ZnO nanowire regions (in the tan color) are in circles of varying densities. (**C**) The maximum z-displacement at each ZnO nanowire region as a function of the number of layers of the CNT, ZnO, and elastomer screen that are between the UV light source and the DEA. The measurement uncertainty is approximately the size of the symbols.

Based on these results, a multilayer consisting of ten active layer DEA with a ZnO nanowire density of 0.25 μg/mm$^2$ was fabricated. As with the evaluation of the single active layer DEA in the Supplementary Materials, the multilayer DEA sheet was clamped into a rigid frame leaving free-standing a square 45mm x 45 mm. The UV light source was placed 20 mm away and shone through a screen having nine 5 mm diameter apertures to evaluate the local actuation of the multilayer DEA. The displacement fields of the DEA for the nine different locations illuminated are shown in Fig. 5.

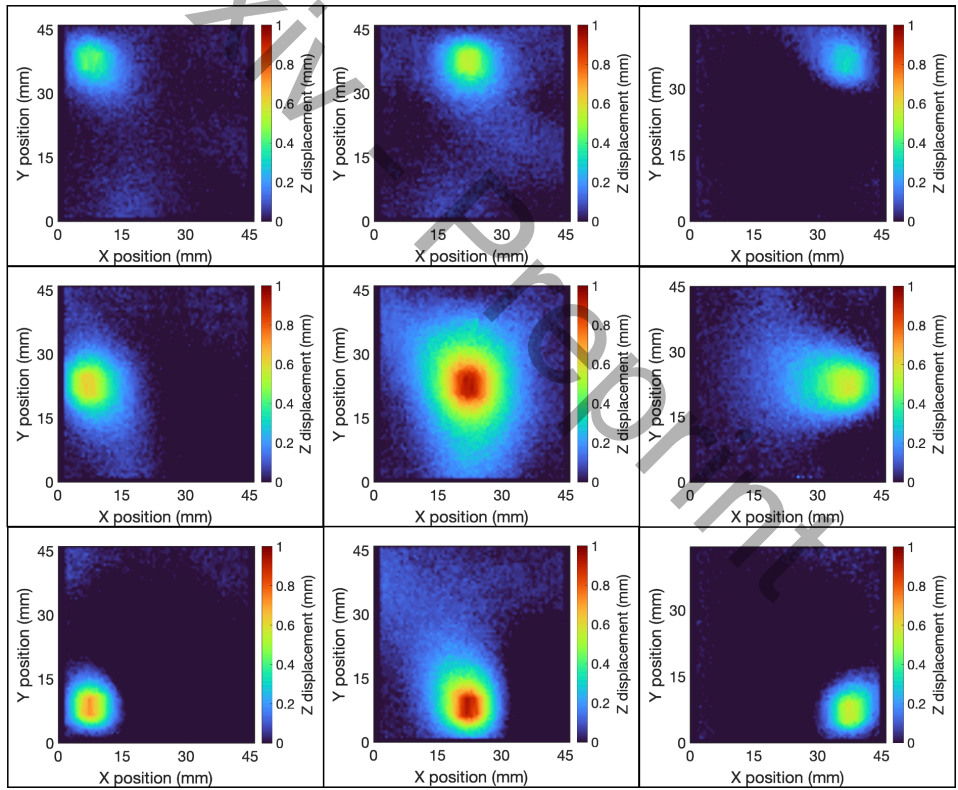

**Fig. 5. Local addressing of multilayer DEA with optically configurable electrodes.** The magnitude of the out-of-plane (z) displacement as a function of position for a ten active layer DEA with a UV light shone at nine different locations multilayer DEA through 5 mm diameter apertures. The DEA is clamped at its edges and free to displace in the z-direction. The maximum out-of-plane deflection is almost 1 mm. The UV beam

itself cannot be detected as its frequency is below the sensitivity of the profilometer detector.

Finite element method (FEM) simulations were performed to compare with these experiments. The case where the UV beam is shining through the center hole into the center of the DEA (position 5 in Fig. 3B) was modeled in Abaqus using a custom-made user defined element for modeling DEAs (*31*), with an individual layer thickness, shear modulus, and relative permittivity of 54 microns, 435 kPa and 8, respectively. An additional set of experiments were performed using a screen with 2.5 mm diameter apertures, and simulations were performed for these as well. The comparison between the experiments and simulations for both aperture sizes can be seen in Fig. 6 and exhibit an excellent quantitative match.

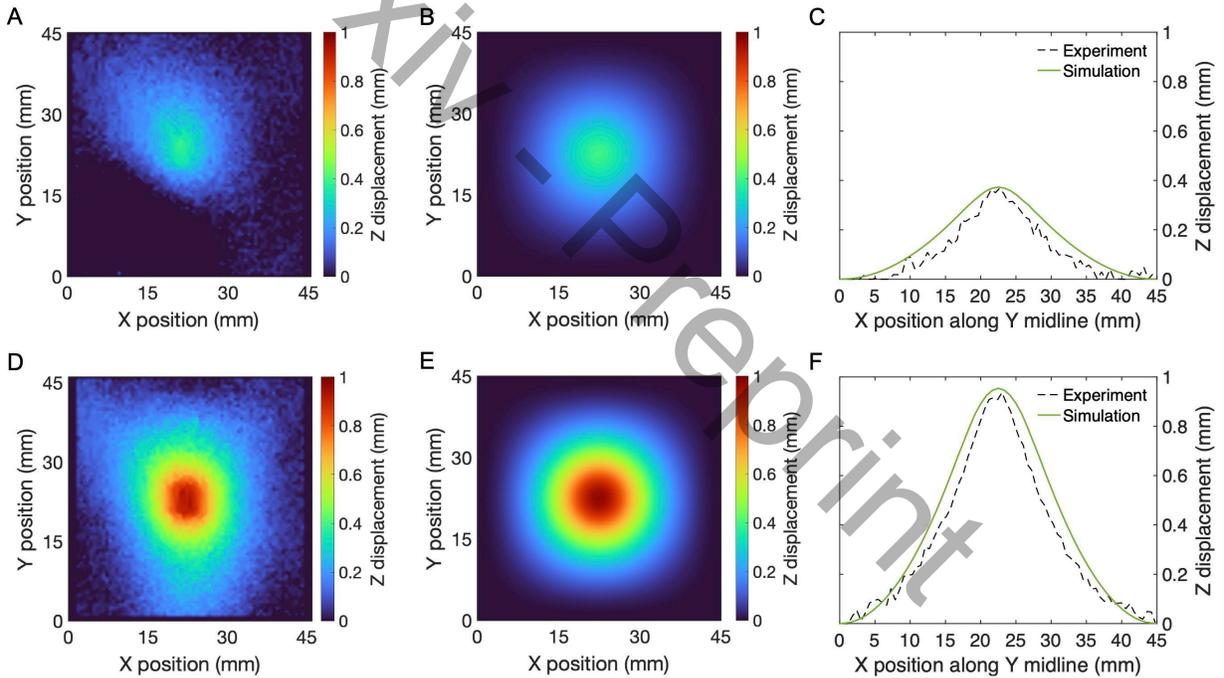

**Fig. 6. Comparison of experimental data and FEM simulations for local DEA actuation.** The measured (**A**) and simulated (**B**) out-of-plane displacements, of a 45mm square, ten active layer DEA illuminated with a 2.5 mm diameter UV beam shone at the center of a DEA clamped around its edges. Electric field per layer 37 V/micron. (**C**) Comparison of the z-displacement along the midline (y = 22.5 mm) for the experiment and simulation in (**A**) and (**B**). (**D**) and (**E**) Corresponding experimental and simulated

out-of-plane deformations for a 5 mm diameter UV beam. (**F**) Comparison of the z-displacement along the midline (y = 22.5 mm) from experiment (**D**) and simulation (**E**).

## 5. DYNAMIC ACTUATION USING "ON-THE-FLY" OPTICAL ADDRESSING

As the UV source is not physically attached to the DEA, it can be moved to produce actuation that follows the UV beam and creates dynamic actuation. Examples of this "on-the-fly" actuation are presented here with images captured from videos shown in the Supplementary Information. A particularly simple example is shown in Video S1, where a UV-controlled ten-layer, 45 mm square DEA was held vertically and clamped at its edges, while a sinusoidal voltage is applied. As the UV-light is translated over the square DEA, the region of actuation is also seen to follow the UV beam. Another example, reproduced in Fig. 7(A-C) from Video S2, is the movement of a ceramic sphere on a horizontal, ten-layer, 45 mm square DEA. Although the UV beam itself cannot be imaged by the camera, its location is revealed by the longer wavelength components that pass through the DEA and appears as the light blueish-green square in the photographs. Fig. 7 (A-C) consists of two successive frames from Video S2 and a superposition of the two. In Fig. 7(A), the sphere is located at the center of the UV beam. Then, when the beam is moved to a new location shown in Fig. 7(B), the sphere rolls to the new beam position. (The rolling action is not captured in the "stills"). The overlay of the two images in Fig. 7(C) shows the movement of the sphere.

The reason the sphere follows the beam is as follows: When the sphere is initially placed on the surface of the DEA with voltage off, it causes the DEA sheet to deflect under the gravitational force on the sphere, and the sphere moves to a position of minimum gravitational plus elastic strain energy in the DEA sheet. When voltage and the UV beam are turned on, the DEA sheet expands biaxially where the UV beam is shone. This local expansion causes the sheet to deform more and changes the location of the global minimum for the potential energy for the sphere. If it can overcome friction, the sphere rolls to this lower energy position. The results of an Abaqus simulation in Fig. 7(D) depict how the location of the minimum total energy changes for three cases. The first is where gravity deforms the DEA surface due to the mass of the DEA sheet alone. The second is where the sphere deforms the DEA sheet but no actuation is induced in the DEA. The third is where the sphere will roll (in this case, toward the center) due to the actuation changing the surface profile of the DEA sheet as well as the gravitational forces. The surface contours in Fig. 7 (E-G) show how the surface contour changes for these three cases. More detailed information on the simulation methodology is described in the Materials and Methods Section.

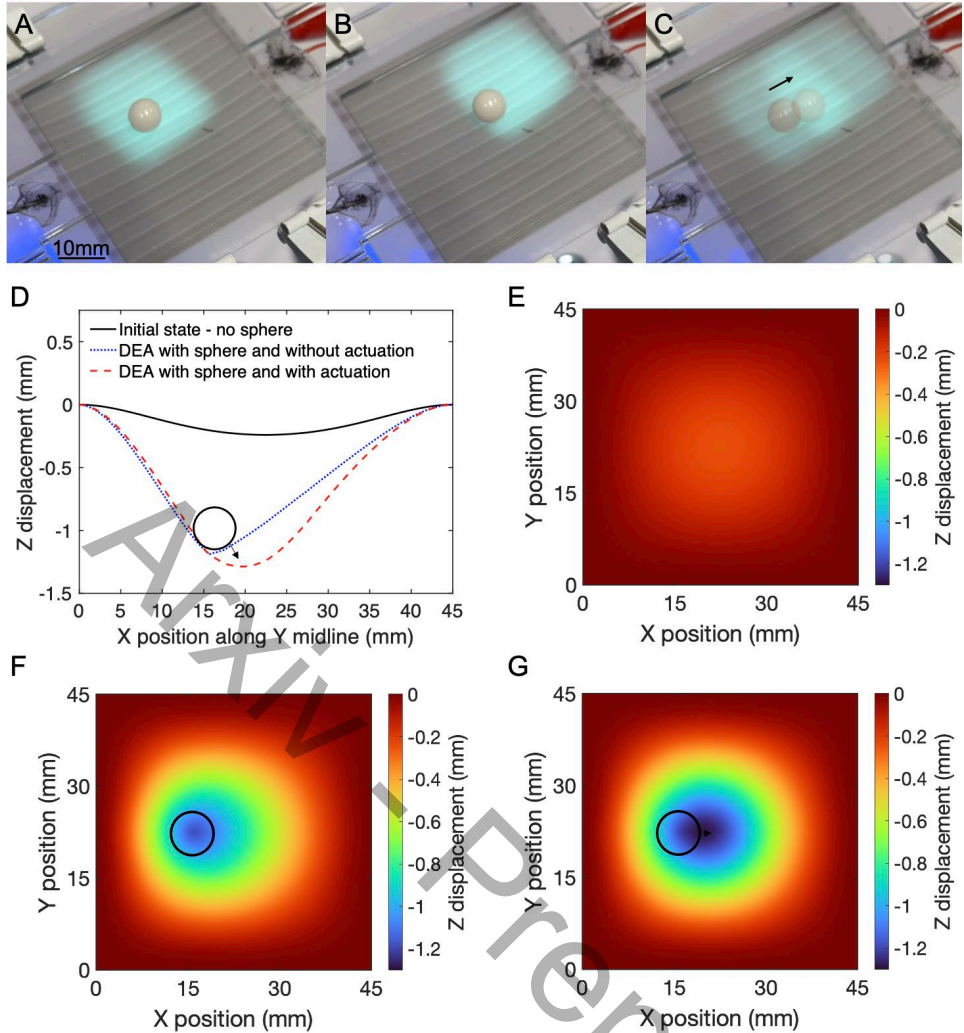

**Fig. 7. UV-beam controlled movement of a ceramic sphere on a horizontal DEA multilayer sheet.** (**A**) Initial position of the sphere and the UV beam. (**B**) The UV beam is translated, modifying the deformation of the DEA sheet and causing the sphere to roll to a new position following the beam. (**C**) Overlay of the two images showing the net movement of the sphere. Although the UV beam itself cannot be imaged, longer blue/green wavelengths from the UV source illuminate the beam's position. The white lines on the DEA appear bright as they correspond to gaps between the CNT interdigitated electrodes where the optical absorption is least. (**D**) Z-displacement along the centerline (y = 22.5 mm) of a UV-controlled DEA sheet from an Abaqus finite element simulation for a scenario similar to that shown in the images of (**A-C**). (**E-G**) Contour maps of the FEM simulated out-of-plane displacements of a ceramic sphere on a UV-controlled DEA (**E**) before a ball is placed on the surface, (**F**) after placing the sphere off-center on the DEA, and (**G**) after actuating a central region of the DEA with UV light. The superimposed black circle indicates the location of the sphere. Electric field per layer of 37 V/micron.

Manuscript Template    Page **14** of **31**

The actuation of a bimorph fin to create a flapping motion is an example of the simpler wiring and remote control afforded by using UV induced photo-conductance electrodes. The bimorph fin is shown schematically in figure 8. It consists of two, back-to-back DEA multilayers joined with a UV-absorbing carbon black layer along the neutral plane and clamped at one end to create a cantilever beam. Three thin PET strips were attached to both sides, stiffening the actuator in the vertical direction so that the bending in response to the biaxial actuation would only occur in a direction perpendicular to the fin. A high voltage was applied at the same time to both multilayers and the multilayer on each side was then alternatively illuminated by a UV source. When the UV source was shone on one side, the multilayer on that side expanded but was constrained by the inactive multilayer on the other side. Consequently, the bimorph bent towards the UV source in response to the UV illumination. When the UV source on the other side was turned on and the initial side turned off, the bending direction reversed. By alternating the side illuminated by the UV source, a flapping motion was created. This flapping-like motion is shown in Fig. 8 (A-C) from individual frames in Video S3. In contrast to previous bimorph flapping fins (*32*), no high voltage switch is needed and only one high voltage supply is needed, the switching is not only non-contact and can be achieved with low voltage UV diodes.

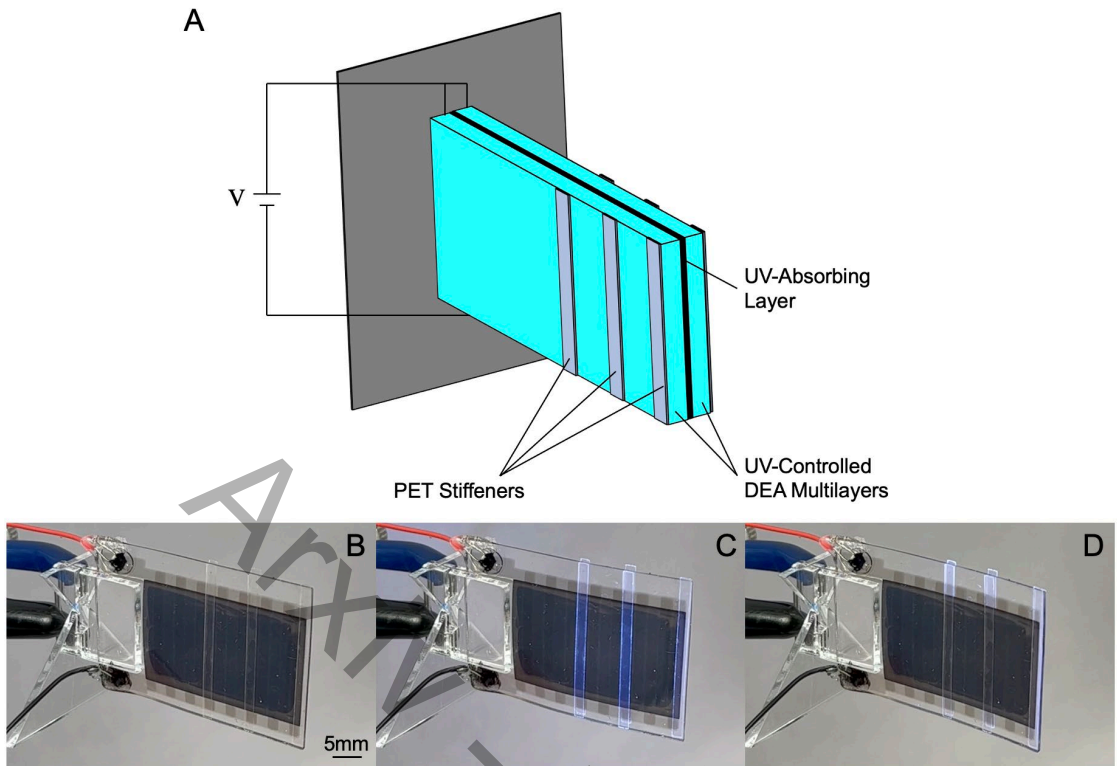

**Fig. 8. UV-controlled bimorph fin actuation.** (**A**) A modified version of the UV-controlled DEA was used, together with an inactive UV-absorbing layer between them along the neutral axis. Silicone release film stiffeners were placed on the front and back surfaces so that the actuation strain is along the fin axis to achieve a fish-like bending. (**B**) The fin in its unactuated state. (**C**) The UV beam illuminates the front surface, causing the front five active layers to expand biaxially and bend. (**D**) The UV beam illuminates the back of the fin, causing the back five active layers to actuate and bend in the opposite direction. The blue, long wavelength emission from the UV source reflected from the silicone stiffeners indicates the side that the UV beam is illuminating.

The preceding examples are all planar DEA configurations, but other geometries are also possible based on the same optically induced conductance of a ZnO layer to create an ad-hoc electrode structure. For instance, in demonstrating the light driven bending behavior shown in Figure 9 and Video A4, the structure is a tightly rolled planar DEA tubular configuration clamped at one end to form a vertical, high aspect tube, Figure 9 (A). This is geometrically

Manuscript Template                                                                        Page **16** of **31**

the same as the rolled actuator configuration used to create a uniaxial force along the rolled actuator axis. When a DC voltage of 2 kV was applied the tube remained vertical and stationary as the net electric field induced expansion is only in the vertical direction. However, when UV light was shone on the side of the rod it bent away from the light as shown in Figure 9 (B). Subsequently, as shown in the series of Figure 9 (C-D), as the direction of illumination was altered, the tube consistently bent away from the light, creating a "light-repelling" structure. The UV induced bending is analogous to that of the bimorph fin. Where the tube is illuminated, the ZnO layers become conducting creating a local expansion of the elastomer in both the hoop and axial directions of the tube. The axial expansion of the tube on one side causes a net bending moment on the tube to bend away from the UV light.

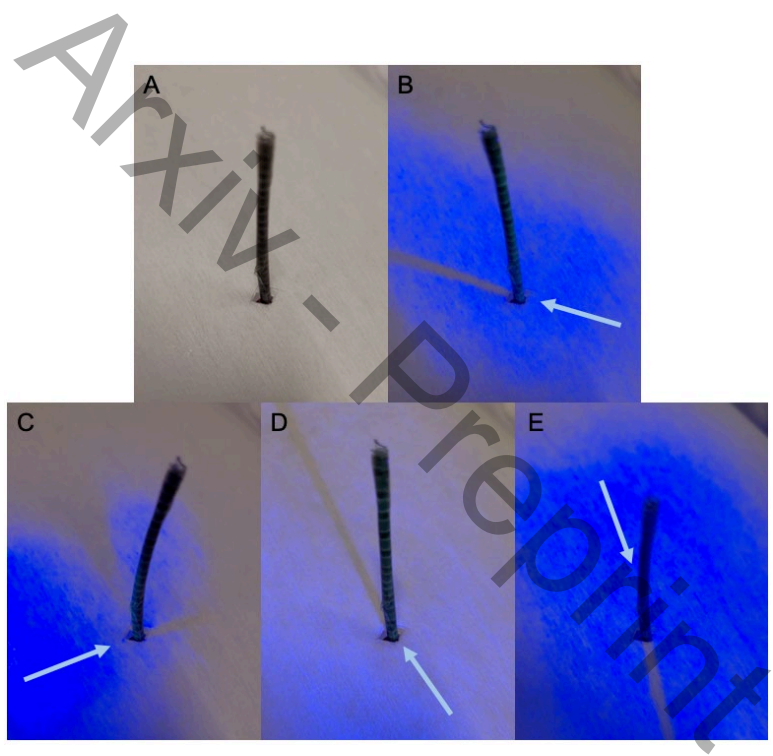

**Figure 9: Light-repelling tube bending.** (A) A DEA tube, formed by tightly winding a 64 x 64 mm square DEA sheet, in its unactuated state and clamped at the bottom in a wooden sheet. (B-E) A series of images showing that the tube bends away from the UV illumination as the direction of the lamp, indicated by the white arrow, is rotated around the tube. The electrical connections, not shown, are beneath the wooden sheet.

## 6. DISCUSSION

The incorporation of photoconductive ZnO nanowire networks embedded in DEAs enables the use of UV light to locally create, define, and configure electrodes for device actuation. As the ZnO nanowires are only conductive when and where they are illuminated, the *effective* electrode structure is not fixed, in contrast with conventional CNT and carbon-black electrodes currently used in DEAs. This provides greater, previously unattainable freedom in the design of dielectric elastomer actuators. It can also decrease the wiring complexity of DEAs, such as for the bimorph fins and arrays of DEAs, by reducing the number of separate high-voltage lines, switches, and control circuitry needed. For instance, to achieve similar results to those shown in Fig. 5 would require an array of nine individual DEAs without the ZnO nanowire electrode architecture and would require ten wires and a method for switching where a voltage is applied. The design presented in this work, on the other hand, requires only three wires, a high-voltage supply, a separate, low voltage circuit for the UV light and a common ground. Also, there are no cross-talk issues unlike the previous works with passive matrices (21). Furthermore, since the actuation occurs where the UV light is shone, new applications can be envisioned such as an elastomeric cover for LCD screens with UV backlight adding a third dimension to the 2D LCD screen using out-of-plane deformation of the DEA.

The ZnO nanowire networks can, in principle, replace conventional CNT and carbon-black electrodes everywhere, although they may be needed as flexible interconnects to the power supplies. The incorporation of the photoconductive electrodes does require additional processing steps, but as shown here, they can be produced by the same stamping process as commonly used to create CNT electrodes in many of today's multilayer DEA technologies. The ZnO nanowire networks do lead to a small increase in stiffness, but this is not expected to adversely affect device performance. They also lead to an increase in optical absorption but as shown here actuation of ten layered multilayer devices is feasible even with a low power UV diode as the excitation signal. The minimum attainable resolution depends on the diameter of the UV beam. One current limitation is the resolution with which the stripes of CNTs can be deposited. The mask cutting techniques currently used limit the CNT stripes to a width of around 1 mm, which in turn limits the minimum width of a region that can be activated by UV light. If the CNT stripes were narrower, the resolution of the activated photoconductive region and newly formed local capacitor could be minimized further. Currently, there are only a few commercial suppliers of UV diodes, and they lack the optics required to focus the beam down to small diameters. However, when they become obtainable, more detailed studies of spatial resolution will be possible. In future, as micro-UV LEDs also become available, it should also be possible to embed them within a DEA further increasing the design and actuation flexibility.

In summary, by implementing a layered ZnO nanowire electrode with a CNT interconnect architecture, a new level of control of complex DEA devices, including fully reprogrammable shape-morphing is possible. Instead of the actuation being limited by predefined electrode locations, this new reprogrammable electrode architecture allows for local level control of DEA actuation not previously attainable. Additionally, it offers the possibility of creating new DEA designs in which the UV beam can be used to move the location of the actuation from one place to another.

## 7. MATERIALS AND METHODS

*Preparation and deposition of ZnO nanowires and CNTs*

ZnO nanowires (ACS Material LLC, SKU: NWZO01A5), with diameters of 50-150 nm and lengths of 5-50 μm, were dispersed in isopropyl alcohol. They were vacuum filtered through a PTFE filter to achieve the desired areal density on the filter (see (*20*) for additional details). Similarly, the CNTs (Carbon Solutions, Inc., P3-SWNT) were first dispersed in deionized water by sonication before being vacuum filtered onto a PTFE filter (see (*21*) for additional details). Both the ZnO nanowires and CNTs were transferred onto the elastomer sheets by stamping through masks, giving precise control over where they were deposited. A silicone release film (Drytac, CRP41082) was used for the mask after been precisely patterned using a Versa laser cutter (VLS 2.30).

*Elastomer materials.*

The elastomer was a UV-curable urethane acrylate (CN 9028) from Sartomer Arkema Group. The elastomer precursor (99.5% CN9028, together with the photoinitiator, 0.5% Diphenyl (2,4,6-trimethylbenzoyl)-phosphine oxide, by mass) were mixed in a centrifugal mixer (Thinky ARE-310) for 25 minutes at 2000 RPMs and defoamed at 2200 RPMs for 5 minutes. It was then deposited as a thin layer by spin coating (Laurell WS-650) onto a PMMA substrate for 1 minute and UV curing each layer in a UV chamber (010034, TECHTONGDA), under nitrogen to prevent oxidation, for 3 minutes.

*UV source.*

The UV light source (DSX033B7W, Shenzhen Deshengxing Technology Co., Ltd.) was an LED having a peak wavelength at 365 nm and an intensity of 70 mW/mm$^2$.

*Fabrication of single and multilayer DEA sheets with ZnO reconfigurable electrodes.*

The fabrication sequence to create a single DEA sheet with optically reconfigurable electrodes is shown in figure 3. CN9028 elastomer was spun coat onto an acrylic substrate at 2000 RPMs for 1 minute and cured under UV light. The CNTs were stamped onto the elastomer through a mask. The mask for the interdigitated CNTs was cut so that one CNT layer has 13 stripes that are 2 mm thick with 2 mm gaps between them and the other CNT layer had only 12 stripes. Then, a continuous layer of ZnO nanowires was stamped directly onto the patterned CNTs and elastomer to form the electrode and electrical connection with the CNT stripes. The ZnO layer completely covered the CNT stripe region. Subsequently, another layer of elastomer was deposited by spin coating and cured before adding another layer of CNT stripes and ZnO nanowires to create both the dielectric and second electrode respectively. As indicated in Fig. 3, the second layer of ZnO nanowires overlapped with first ZnO nanowire layer, whereas the second layer of CNT stripes was offset from the first layer of CNT stripes so that there was no overlap between them. Lastly, a final elastomer layer was spun coat onto the top at a higher rate of 3000 RPMs to make a thinner top elastomer layer. The resulting difference in thickness has the effect of shifting the neutral axis and consequently biasing the direction that the DEA moves out of plane when a voltage and UV illumination are applied. Carbon grease and insulated tinned copper wires were connected to the ground and HV leads on the sample to allow for the application of voltage.

To prepare the multilayer DEA used for the testing in figure 5, the fabrication process for a single DEA layer was repeated and the layers stacked to construct the multilayer. The only difference being that a higher spin coating speed of 3000 RPMs was used for each layer instead, and then the final elastomer layer was deposited at a rate of 4000 RPMs. Once again, the thinner outer layer was used to bias the bending direction when the DEA deflected out of plane.

*Fabrication of the light-bending tube*

The construction of the soft tube is made with an almost identical electrode structure as that shown in Figure 3.4 and described in the previous paragragh, but with the high voltage and ground CNT connection adjacent to each other on the same side and with 2 cm long CNT strips. Additionally, as opposed to using 1.2 mL of the CNT solution for the electrodes a higher concentration (6.0 mL of the CNT solution) in order to create a more UV absorbing layer.

*Deflection measurements.*

Local deflections of the DEA sheets were measured using a non-contact laser profilometer (MTI ProTrak, PT-G 60-40-58) operating at 660 nm.

*Abaqus FEM Simulation: Moving a ceramic sphere on UV-controlled DEA*

The deformation of a square 45 mm DEA sheet and rigidly clamped on four edges was modeled using Abaqus. Prior to applying any voltage, gravity was introduced into the model, creating a sagging of the sheet (Fig. 8E). Using a dynamic, implicit step method in Abaqus, a dense sphere was then dropped under gravity slightly off-center from just above the surface of the DEA, contacted the DEA sheet, and then came to rest on the DEA surface. A top view of the DEA sheet showing the outline of the location of the sphere and the displacement of the DEA due only to gravity and the contact of the sphere is depicted in Fig. 8F.

To actuate the DEA, a static step was implemented to make use of the static user defined element (*31*), with the sphere held in position. The DEA was actuated directly at its center by prescribing a circular region (defined by the diameter of the light beam) where the voltage is now present. The surface profile after this final step is shown in Fig. 8G and indicates that the sphere can be expected to roll to the center of the DEA.

**SUPPLEMENTARY MATERIALS**

**This PDF file includes:**

Supplementary Methods
Figs. S1 to S3

**Other Supplementary Material for this manuscript includes the following:**

Videos S1 to S4

24. S. P. Lacour, H. Prahlad, R. Pelrine, S. Wagner, Mechatronic system of dielectric elastomer actuators addressed by thin film photoconductors on plastic. *Sensors and Actuators A: Physical*, **111(2-3)**, 288-292 (2004).

25. V. Bacheva, A. Firouzeh, E. Leroy, A. Balciunaite, D. Davila, I. Gabay, F. Paratore, M. Bercovici, H. Shea, G. Kaigala, Dynamic control of high-voltage actuator arrays by light-pattern projection on photoconductive switches. *Microsystems & Nanoengineering*, **9(1)**, 59 (2023).

26. S. P. Lacour, S. Wagner, H. Prahlad, R. Pelrine, High voltage photoconductive switches of amorphous silicon for electroactive polymer actuators. *Journal of non-crystalline solids*, **338**, 736-739 (2004).

27. V. Srikant, D. R. Clarke, On the optical band gap of zinc oxide. *Journal of Applied Physics*, **83(10)**, 5447-5451 (1998).

28. D. A. Melnick, Zinc oxide photoconduction, an oxygen adsorption process. *The Journal of Chemical Physics*, **26(5)**, 1136-1146 (1957).

29. J. R. Collins, D. G. Thomas, Photoconduction and surface effects with zinc oxide crystals. *Physical Review*, **112(2)**, 388 (1958).

30. A. N. Gent, A new constitutive relation for rubber. *Rubber chemistry and technology*, **69(1)**, 59-61 (1996).

31. E. Hajiesmaili, D. R. Clarke, Dielectric elastomer actuators. *Journal of Applied Physics*, **129(15)**, 151102 (2021).

32. F. Berlinger, M. Duduta, H. Gloria, D. R. Clarke, R. Nagpal, R. Wood, A modular dielectric elastomer actuator to drive miniature autonomous underwater vehicles. *2018 IEEE International Conference on Robotics and Automation (ICRA)* (IEEE, 2018) pp. 3429-3435.
**ACKNOWLEDGMENTS:** The authors are grateful to Professor Katia Bertoldi and David Farrell for helpful discussions regarding finite element computations.

**FUNDING:** This work was supported by the National Science Foundation through the Harvard
Manuscript Template    Page **24** of **31**

University Materials Research Science and Engineering Center DMR-2011754. GD is also grateful to the NSF for an NSF Graduate Research Fellowship.

**Author contributions:**

    Conceptualization: EH, GD, DRC
    Methodology: EH, GD, DRC
    Testing: EH, GD
    Visualization: EH, GD, DRC
    Supervision: DRC
    Writing – original draft: GD
    Writing – review & editing: GD, EH, DRC

**COMPETING INTERESTS:** The authors declare that they have no competing interests. A provisional patent describing the use of photoconductive ZnO nanowires in dielectric elastomer actuators has been filed.

**DATA AND MATERIALS AVAILABILITY:** All data needed to evaluate the conclusions of the paper are available in the paper or the Supplementary Materials.

# SUPPLEMENTARY MATERIALS FOR

## OPTICALLY RECONFIGURABLE ELECTRODES FOR DIELECTRIC ELASTOMER ACTUATORS

Gino Domel,[1]*[†] Ehsan Hajiesmaili,[1][†][‡] David R. Clarke[1]

[1]John A. Paulson School of Engineering and Applied Sciences, Harvard University; Cambridge, MA, USA.

[†] These authors contributed equally to this work.

[‡] Now at Meta Reality Labs, Redmond, WA 98052.

*Corresponding author. Email: clarke@seas.harvard.edu

## Supplementary Methods

### 1. Effect of ZnO nanowires on elastic stiffness

Tensile test samples were fabricated based on the ASTM D412 standard. Each consisted of six layers of elastomer encapsulating between them a total of five layers of ZnO nanowires. The urethane acrylate elastomer precursor (CN9028) was deposited in thin layers by spin coating at a speed of 3000 RPM and then UV cured in a UV chamber, and the ZnO nanowires were deposited by stamp transferring from a PTFE filter. Three ZnO nanowire areal densities were evaluated: 0.25, 0.5 and 1.0 μg/mm$^2$.

In the tests, dog-bone shaped samples were loaded in tension, while a force sensor measured the resulting force. The stress as a function of stretch is shown in Fig. S1. Fitting the stress-stretch curves to the Gent model (*30*) of deformation showed a range of 417-467 kPa for the shear moduli of the samples with no ZnO nanowires and varying ZnO nanowire densities, indicating that the ZnO nanowire layers have only a minor effect on the stiffness of the elastomer for even the highest ZnO density explored.

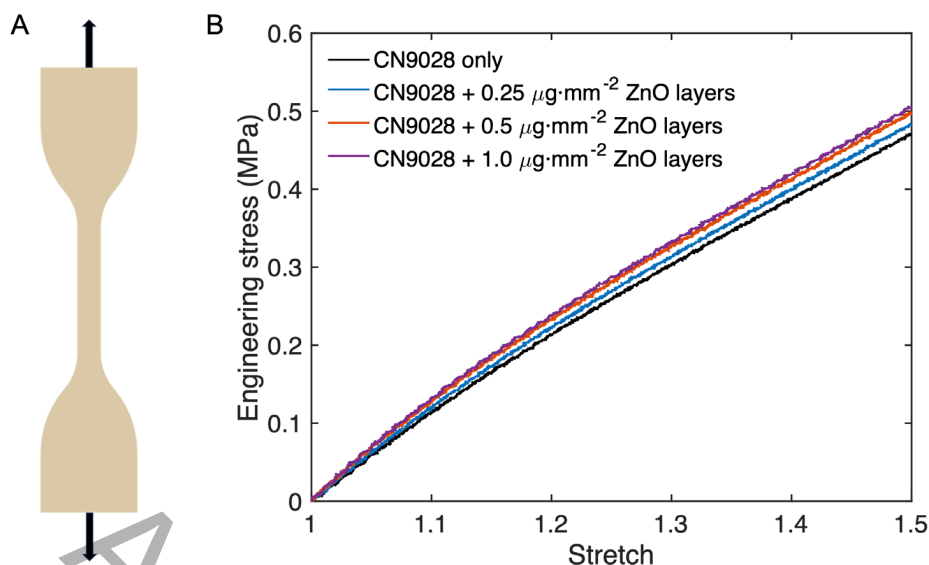

**Fig. S1. Stress-stretch curves for a CN9028 elastomer containing embedded ZnO nanowires.** (**A**) Dog-bone samples were used to evaluate the change of stiffness in the elastomer by embedding ZnO nanowires. (**B**) Based on curve fitting to the Gent model, the shear moduli were 417 kPa, 435 kPa, 458 kPa, and 467 kPa for the elastomer with the density of ZnO nanowires indicated, indicating a modest increase in moduli with increasing ZnO areal density.

## 2. Effect of strain on embedded ZnO nanowire photoconductance

For these tests, the samples were comprised of two layers of elastomer encapsulating one layer of CNTs and ZnO nanowires with the same geometry as shown in Figure 2A. The elastomer precursor was deposited onto a PMMA substrate by spin coating at 1500 RPM for 1 minute and then UV cured. Both the CNTs and the ZnO nanowires were deposited as thin layers by stamping through a mask to give precise control over where they were deposited. Each sample had two physically isolated CNT regions, separated by 1 mm, electrically bridged only by the elastomer layer with ZnO nanowires. The length of the ZnO nanowire bridge was 3 mm, with 1 mm of overlap with each of the CNT regions to ensure electrical connection. The width of the ZnO nanowire bridge and the CNT regions was 3 mm, the same as the dog-bone.

The samples were loaded into the same custom tensile testing apparatus as used in the preceding section, with one end of the sample connected to the high voltage (HV) source (Trek 610E, Advanced Energy), and the other to the ground with a current meter in series. A UV light source (DSX033B7W, Shenzhen Deshengxing Technology Co., Ltd.) with a peak wavelength of 365 nm and light intensity of 70 mW/mm$^2$ was placed 20 mm from the sample with the center of the UV light illuminating the ZnO nanowire bridge. With 2 kV applied (though automatically modulated if a maximum current is reached), current measurements were recorded with the UV light off for the first 5 seconds, on for 10 seconds, and then off again as indicated by the white and blue shaded regions in the graphs in Fig. 3B-D. Current measurements were taken for each sample in its unstretched state and then at incremental stretches of 0.05 from 1.05 up to a maximum of 1.5.

## 3. Local addressing of a single layer DEA membrane

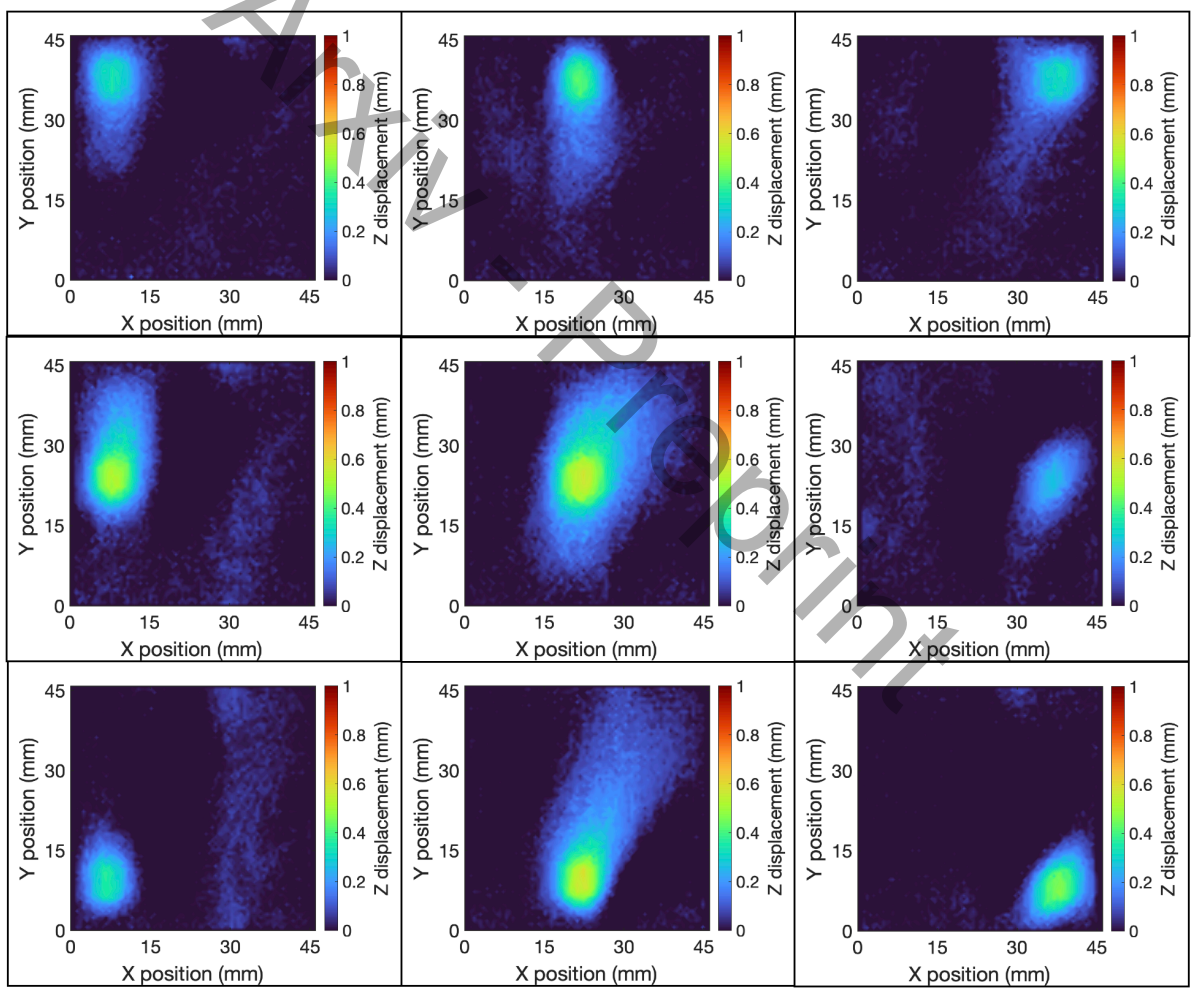

**Fig. S2**. **Local addressing of a single layer DEA with an optically reconfigurable electrode.** The out-of-plane (z) displacement as a function of position for a single layer DEA with a UV light shone at nine different locations on the 45 mm x 45 mm

DEA through a screen with 5 mm diameter apertures. The DEA is clamped at its edges and free to displace in the z-direction. Voltage 2 kV.

## 4. UV absorption measurements

Optical absorption measurements of both the CN9028 elastomer and elastomer sheets containing different areal densities of embedded ZnO nanowires were recorded using a HORIBA Duetta spectrometer.

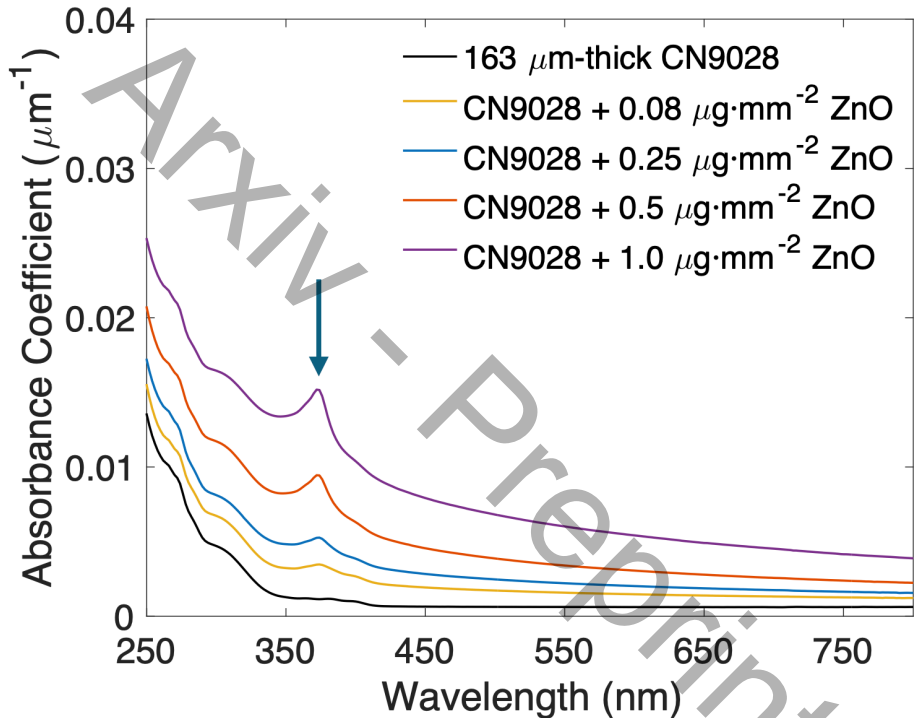

**Fig. S3.** Absorbance coefficient as a function of wavelength for a 163 μm thick sample of CN9028 and for 163 μm thick CN9028 elastomers with the indicated areal densities of embedded ZnO nanowires. The arrow indicates the wavelength corresponding to the absorption of the ZnO nanowires. The absorption peak increases as the ZnO density increases.

## 5. Fabrication of bimorph fin

The bimorph fin was built as a multilayer from individual DEA layers following the procedure described in the main text for the optically reconfigurable multilayer of this work. The same CN9028-based elastomer precursor was used and deposited in thin layers by spin coating onto a PMMA substrate at 3000 RPMs. CNTs for the interconnects were deposited by stamp transferring through a mask, though this mask only had 7 stripes that were 20 mm x 2 mm with a gap between them of 2 mm. A 30 mm x 20 mm ZnO nanowire layer was then deposited by stamping the ZnO-coated filter directly on top of the CNTs. After depositing another elastomer layer, the next layer of CNT stripes was deposited with an offset from the previous layers to prevent any overlap between CNT layers. Another ZnO nanowire layer was deposited on top of this layer and perfectly overlapped with the previous ZnO nanowire layer.

This process was repeated with alternating offset CNT layers until 6 electrode layers and 7 elastomer layers were deposited, creating five stacked capacitors. A layer of dense carbon black was deposited using the same method filtration and mask methods and with the same dimensions as the ZnO nanowire layers. As previously described, this carbon black layer absorbed UV when the light was shone on the device. Seven more elastomer layers and six more electrode layers were then deposited, creating the second half of the fin. Again, carbon grease and insulated tinned copper wires were connected to the ground and HV leads on the fin to allow for the application of voltage. Additionally, 6 strips of silicone release film that are 27 mm x 1 mm are placed on the two large sides of the fin with 3 on each side. These silicone release film strips partially constrain the expansion of the fin along their length, encouraging the fin to bend in a fish-like motion.

## Supplementary Videos

**Video S1**. **Dynamic UV beam controlled movement of a vertically clamped DEA multilayer sheet**. A voltage of 2 kV with a frequency of 1 Hz is applied as the UV beam location is moved around, causing the DEA to locally deform. Although the UV beam itself cannot be imaged, longer wavelengths from the UV source illuminate the beam's position.

**Video S2**. **Dynamic UV beam controlled movement of a ceramic sphere on a horizontal DEA multilayer sheet at 4x speed.** A voltage of 2 kV is applied to a square DEA sheet with a ceramic sphere resting on its surface. As the UV beam, which is shining on the DEA from below, is moved, the DEA locally deforms at the location of the beam, causing the sphere to roll following the beam. Although the UV beam itself cannot be imaged, longer wavelengths from the UV source illuminate the beam's position.

**Video S3**. **UV-controlled bimorph fin.** The fin is clamped at one end, and 1.6 kV AC at 1Hz is applied to the DEA fin. A UV lamp is located on each side of the fin. Both UV sources are turned on for one second and off for one second but out of phase with each other so that only one UV light source is on at any time. Thus, only one side of the fin is illuminated at a time, creating an alternating biaxial expansion and bending motion that mimics a fin. The blue, long wavelength emission from the UV source reflected from the silicone release film stiffeners indicates the side that the UV beam is illuminating.

**Video S4. UV light repelling tube**. A tubular DEA, clamped at one end, bends away from the source of the UV illumination as the direction of the UV lamp is rotated around the axis of the tube. A voltage of 2 kV is applied through the wall thickness of the tube. When no voltage is applied the tube is vertical and does not respond to the UV illumination.

Manuscript Template Page **31** of **31**